# Behaviour of molecular hydrogen emission in three solar flares

Sargam M. Mulay,[1]⋆ Lyndsay Fletcher,[1,2]† Hugh Hudson[1]‡ Nicolas Labrosse[1]§

[1] *School of Physics & Astronomy, University of Glasgow, G12 8QQ, Glasgow, UK*
[2] *Rosseland Centre for Solar Physics, University of Oslo, P.O.Box 1029 Blindern, NO-0315 Oslo, Norway*

**ABSTRACT**

We have systematically investigated ultraviolet (UV) emission from molecular hydrogen ($H_2$) using the Interface Region Imaging Spectrometer (IRIS), during three X-ray flares of C5.1, C9.7 and X1.0 classes on Oct. 25, 2014. Significant emission from five $H_2$ spectral lines appeared in the flare ribbons, interpreted as photo-excitation (fluorescence) due to the absorption of UV radiation from two Si IV spectral lines. The $H_2$ profiles were broad and consisted of two non-stationary components in red and in the blue wings of the line in addition to the stationary component. The red (blue) wing components showed small red-shifts (blue-shifts) of ∼5–15 km s$^{-1}$ (∼5–10 km s$^{-1}$). The nonthermal velocities were found to be ∼5–15 km s$^{-1}$. The interrelation between intensities of $H_2$ lines and their branching ratios confirmed that $H_2$ emission formed under optically thin plasma conditions. There is a strong spatial and temporal correlation between Si IV and $H_2$ emission, but the $H_2$ emission is more extended and diffuse, further suggesting $H_2$ fluorescence, and - by analogy with flare "back-warming"- providing a means to estimate the depth from which the $H_2$ emission originates. We find that this is 1871±157 km and 1207±112 km below the source of the Si IV emission, in two different ribbon locations.

## 1 INTRODUCTION

Molecular hydrogen ($H_2$), formed in the cool chromosphere, emits line radiation in the ultraviolet (UV) when fluoresced by strong UV emission formed in higher layers (Jordan et al. 1977, 1978; Bartoe et al. 1979; Labrosse et al. 2007). Fluoresced $H_2$ lines have recently been reported by Mulay & Fletcher (2021) in solar-flare ribbons from the *Interface Region Imaging Spectrograph* (IRIS) spacecraft (De Pontieu et al. 2014). In flares, the emission lines providing the UV exciter wavelengths can brighten by factors of 100 over their quiet Sun intensities, giving a $H_2$ signal that can be studied in some detail. Flares result in strong heating of the chromosphere, but clearly, some $H_2$ survives, presumably in the coolest part of the chromosphere – the temperature minimum region (TMR), and the properties of these molecular lines can give us a direct insight into the condition of the plasma in which they are formed. Conditions in the TMR in flares are of interest since there is a lack of understanding of whether, and by what means, the TMR could be heated during a flare.

In a previous paper (Mulay & Fletcher 2021), we reported on the discovery with IRIS of fluoresced $H_2$ UV emission in a flare and described the basic properties of the $H_2$ line at 1333.797 Å. In this line, the UV exciter is emission from Si IV at 1402.77 Å, nominally emitted at log $T$ [K] ≈ 4.8 (63100 K). We found the Doppler shift of this $H_2$ line to be consistent with zero within observational errors, and the non-thermal line width was around 10 km s$^{-1}$. There was a strong linear correlation of the intensity with the 1402.77 Å intensity, but with two different gradients in the linear relationship in two different flare ribbons. From the intensity ratio of the two Si IV lines at 1393.76 Å and 1402.77 Å (Mathioudakis et al. 1999; Gontikakis & Vial 2018; Tripathi et al. 2020; Zhou et al. 2022), we concluded that the opacity at the Si IV exciter wavelengths was different in the two ribbons, indicating different chromospheric conditions that could influence the formation or transfer of the $H_2$ line. Fluoresced $H_2$ emission was also reported in a number of microflares by Innes (2008).

In this paper, we expand our study of fluoresced $H_2$ emission, using the series of flares that took place in active region NOAA #12192 on October 25, 2014, culminating in the X1.0 event SOL2014-10-25T15:04. This region was the largest recorded in Solar Cycle 24, and produced 6 X-class flares and numerous M-class flares between 18$^{th}$ and 29$^{th}$ October 2014 (Chen et al. 2015). Unusually for such a large region, none of these X-flares resulted in a coronal mass ejection. The magnetic configuration and evolution of region NOAA #12192 have been the subject of extensive study; in particular, the events of October 25, 2014, prior to the X1.0 flare have been examined by Bamba et al. (2017). Kowalski et al. (2019) investigated the emission from cool metallic lines and continuum in the near UV (NUV) channel of IRIS, in umbral brightenings of what we will refer to in this paper as "Ribbon 3" (R3). Ashfield et al. (2022) also looked at this region of the flare, applying a technique known as the "UV footpoint calorimeter" to determine the energy flux into flare ribbons. Unfortunately, at the times and locations studied by these authors, the counts are too low to fit the $H_2$ lines.

In this paper, we focus on the $H_2$ flare ribbon emission, produced by much cooler plasma, along with its exciting Si IV lines. Several H$_2$ lines have been detected, including two pairs of lines in which the transition upper level is the same. This allows us to measure the branching ratio and compare it with the theoretical branching ratio to test for opacity effects. We also examine the spatial distribution of the fluoresced H$_2$ emission compared to the fluorescing Si IV line, to look for the "core-halo" effect that might be expected from this excitation mechanism, in which the Si IV radiation spreading isotropically excites a larger patch of H$_2$ in the lower atmosphere.

⋆ E-mail: Sargam.Mulay@glasgow.ac.uk
† E-mail: Lyndsay.Fletcher@glasgow.ac.uk
‡ E-mail: Hugh.Hudson@glasgow.ac.uk
§ E-mail: Nicolas.Labrosse@glasgow.ac.uk

The paper is organized as follows. Section 2 provides the overview of the UV spectroscopic and imaging observations. A detailed investigation of three X-ray flares along with their UV spectral signatures is discussed in Section 3. In Section 4, we study spectral profiles, deriving Doppler and nonthermal velocities for H$_2$ lines. In Section 5, we apply a simple geometric model developed for flare radiative back-warming to find the depth of the H$_2$ emission. We discuss and summarize our results in Section 6.

## 2 OBSERVATIONS AND DATA ANALYSIS

### 2.1 Overview

Three X-ray flares of C5.1, C9.7 and X1.0 classes occurred in active region #12192 (S12 W35[1]) on Oct. 25, 2014 (SOL2014-10-25T15:04 (Leibacher et al. 2010)). The region had a complex $\beta\gamma\delta$ magnetic configuration. The X-ray fluxes were observed by the *Geostationary Operational Environment Satellite* (GOES-15). Figure 1 panels (a-b) provides the GOES X-ray fluxes of the flares in two channels, 0.5-4.0 Å (orange curve) and 1.0-8.0 Å (black curve) and their derivatives, respectively. The X-ray activity from 15 to 18 UT shows a systematic increase in the GOES flare class and duration of the flares. The GOES start and end timings of the flares are shown by grey shaded areas, whereas the green shaded areas indicate the time slots where emission from molecular hydrogen, H$_2$ was observed at the flare ribbons by IRIS (please note, even when this emission is observed it is very often not strong enough to fit the line profile). Table 1 provides the GOES X-ray flare start, peak, and end timings and the start and end times for the H$_2$ emission during the flares where we performed line fittings.

The X-ray fluxes in four channels observed by the *Reuven Ramaty High Energy Solar Spectroscopic Imager* (RHESSI) satellite (Lin et al. 2002) are plotted in Fig. 1 panel (c), showing the eclipse data gaps from 15:35 to 15:55 UT and 17:07 to 17:31 UT. Where GOES data are available, the GOES time-derivative curves show peaks very close in time to the highest energy RHESSI light curves, as would be expected from the Neupert effect (Neupert 1968; Dennis & Zarro 1993).

The chromospheric signatures of flares were observed in the 1600 Å images at the peak time of GOES X-ray flares (see Fig. 1, panel d−f) obtained from the *Atmospheric Imaging Assembly* instrument (AIA; Lemen et al. 2012) on board the *Solar Dynamics Observatory* (SDO; Pesnell et al. 2012). We used the Joint Science Operations Center (JSOC)[2] to download level 1 data from the AIA 1600 Å channel. The initial processing of the data was carried out using the standard AIA routines available in the Solarsoft libraries (SSW; Freeland & Handy 1998). The 1600 Å images showed that the three flares occurred at very similar locations between two sunspots (see Fig. 1, panel (e))

The *Slit-Jaw Imager* (SJI) onboard IRIS observed active region #12192 from 14:58:28 to 18:00:47 UT in the C II 1330 Å, and Mg II 2796, 2832 Å windows with a cadence of 16 sec and resolution of 0.33″. Table 2 summarizes the IRIS observation details. The field-of-view (FOV) captured by SJI is shown in Fig. 1, panel (d) as a black-boxed region, and a zoomed-in view of this region is shown in C II 1330 Å images (see Fig. 1, panels g−i). The deep red patches in 1600 Å and C II images are three flare ribbons indicated by black arrows. We named these Ribbon 1 ("R1"), Ribbon 2 ("R2"), and Ribbon 3 ("R3") in order of their occurrence during flaring activity. The IRIS SJI C II images showed that emission at R1 was observed during all three flares, whereas emission at R2 and R3 was observed only during the X1.0 flare.

Simultaneously, the IRIS *spectrograph* (SG) captured the full spectra in three wavelength ranges, 1331.69−1358.32 Å, 1380.77−1406.70 Å, and 2783.32−2834.96 Å. IRIS level 2 data downloaded from the IRIS database[3] was processed using the standard Solarsoft routines[4]. We corrected the data for orbital variation and the absolute wavelength calibration was carried out using the strong photospheric O I 1355.6 Å line which is assumed to be unaffected by Doppler shifts during a flare. The spectrograph slit was positioned horizontally (with a rotational angle of +90°) in a sit-and-stare mode (2040 slit exposures were made) crossing the three flare ribbons and the umbra-penumbra of a sunspot. The slit position is indicated in Fig. 1, panel (d) as a horizontal black line and is also shown in C II 1330 Å images (panels g−i). The IRIS observations started at 14:58:28 UT, two minutes before the C5.1 flare started, and ended at 18:00:47 UT. The spectrograph slit captured emission from three flare ribbons, and we obtained spectra for these ribbons in the green boxed regions which are shown in C II images (see Fig. 1, panels g−i) (R1: X-axis = 430″ to 450″ and Y-axis = -320″, R2: X-axis = 410″ to 430″ and Y-axis = -320″, R3: X-axis = 355″ to 375″ and Y-axis = -320″).

We refer to the comprehensive atlas of UV lines reported in Sandlin et al. (1986) and the atlas provided by SESAM molecular spectroscopy database[5] to identify H$_2$ lines. Our analysis is focused on two Si IV lines at 1393.76 and 1402.77 Å in addition to five unblended H$_2$ lines at 1333.481, 1333.797, 1338.565, 1342.257 and 1398.954 Å. We observed that the C II and Si IV line profiles were very broad. So it was difficult to identify the presence of other H$_2$ lines (1334.486, 1334.501, 1335.549, 1335.581, 1393.451, 1393.719, 1393.732, 1393.901, 1395.199, 1402.648, and 1403.381 Å) that are close to C II 1335.71 Å, Si IV 1393.76 Å and Si IV 1402.77 Å lines. In addition, we observed three blended H$_2$ lines − 1) H$_2$ line at 1346.911 Å which is blended with Si II at 1346.878 Å, 2) H$_2$ line at 1396.225 Å which is blended with S I at 1396.108 Å, 3) H$_2$ line at 1352.504 Å which is blended with Si II at 1352.605 Å. It was difficult to separate these H$_2$ lines from their blends. So we could not use these lines for further analysis. Therefore the only H$_2$ lines we could use for our analysis are the five unblended H$_2$ lines at 1333.481, 1333.797, 1338.565, 1342.257, and 1398.954 Å.

Figure 2, panel (a) shows the GOES X-ray fluxes for three flares (same as Fig. 1, panel (a)) along with the light curves for AIA 1600 Å and SJI C II 1330 Å in panel (b). They are obtained by averaging the intensities over pixels covered by the FOV shown in Fig. 1, panels (d) and (g) respectively. Slow variations in the UV light curves are seen outside the impulsive phase times, similar to GOES, but much more

---

[1] https://www.solarmonitor.org/?date=20141025
[2] http://jsoc.stanford.edu/
[3] https://iris.lmsal.com/data.html
[4] https://iris.lmsal.com/itn26/
[5] http://sesam.obspm.fr/

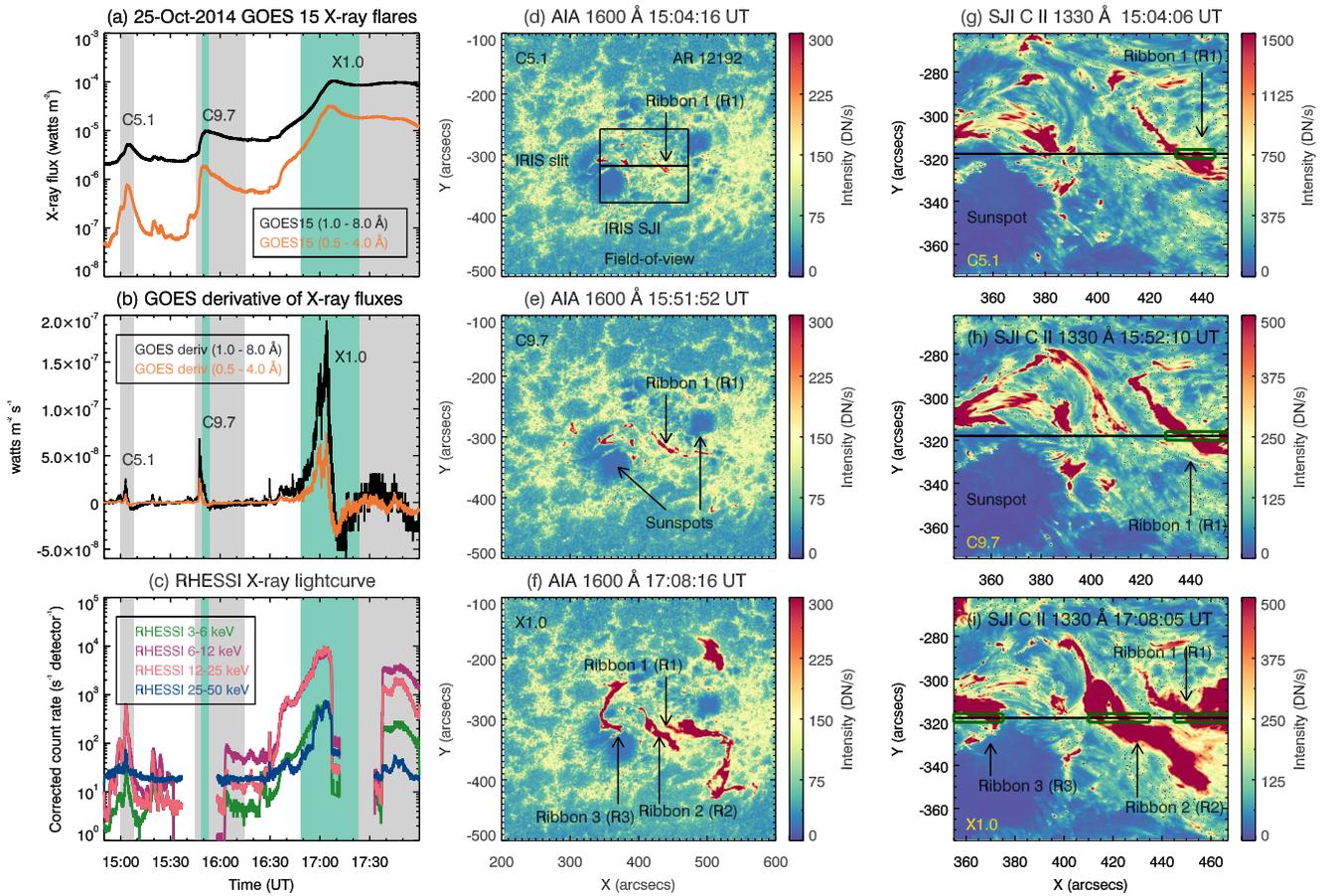

**Figure 1.** Panel (a): X-ray fluxes for three flares − C5.1, C9.7, and X1.0 − observed on Oct. 25, 2014, by two GOES-15 channels, 1.0-8.0 Å (black curve) and 0.5-4.0 Å (orange curve), and Panel (b): the time derivative of X-ray fluxes. Panel (c): The soft and hard X-ray fluxes in various energy ranges observed by the RHESSI satellite. The grey shaded areas indicate the duration of each GOES flare, and the green shaded areas indicate the time slots where we performed line fitting for $H_2$ emission during the flares. Panel (d-f): AIA 1600 Å images of the active region (AR) #12192 at the GOES X-ray peak times of three flares. The IRIS SJI field-of-view (FOV) is indicated by a black box, and the black horizontal line indicates the IRIS spectrograph slit position (with a rotational angle of +90°). Panel (g-i): A zoomed FOV of IRIS SJI is shown in the C II 1330 Å window images that are obtained at the GOES X-ray peak times of each flare. The deep red patches indicate emission from three flare ribbons (R1, R2, and R3). The spectra in these ribbons (see Fig. 3) were obtained for regions that are shown by green boxes and black arrows (R1: X-axis = 430″ to 450″ and Y-axis = -320″, R2: X-axis = 410″ to 430″ and Y-axis = -320″, R3: X-axis = 355″ to 375″ and Y-axis = -320″).

**Table 1.** GOES X-ray flare timings and timings for the $H_2$ emission observed at various ribbon locations

| | GOES X-ray flare timings | | | | $H_2$ emission (Ribbon 1) | | $H_2$ emission (Ribbon 2) | | $H_2$ emission (Ribbon 3) | |
|---|---|---|---|---|---|---|---|---|---|---|
| X-ray class | Start (UT) | Peak (UT) | End (UT) | Flare duration (hh:mm:ss) | Start-End time (UT) | $H_2$ duration (hh:mm:ss) | Start-End time (UT) | $H_2$ duration (hh:mm:ss) | Start-End time (UT) | $H_2$ duration (hh:mm:ss) |
| C5.1 | 15:00 | 15:04 | 15:08 | 00:08:00 | − | − | − | − | − | − |
| C9.7 | 15:44 | 15:52 | 16:15 | 00:31:00 | 15:49:16−15:50:52 | 00:01:36 | − | − | − | − |
| X1.0 | 16:55 | 17:08 | 18:11 | 01:16:00 | 16:48:07−16:58:28 | 00:10:21 | 17:03:23−17:06:41 | 00:03:18 | 16:50:25−17:21:48 | 00:31:23 |

**Notes -** GOES flare timings given in columns 2−4 are according to the X-ray fluxes observed in the 1.0−8.0 Å channel.

impulsive behaviour is seen during the flares. The temporal evolution of flare ribbon emission in Si IV and $H_2$ lines is studied along each slit position. We created spectral images for Si IV (panel c) and $H_2$ (panel d) lines by integrating the total intensity over the Si IV line at 1402.77 Å, summing from 1402.59 to 1403.41 Å and over the $H_2$ line at 1333.797 Å, summing from 1333.66 to 1333.97 Å. The emission from flare ribbons is seen as a white patchy structure and is shown by white arrows. The $H_2$ emission maps the Si IV very well in space and time but has a more diffuse appearance in places, extending further in the y-direction. Note that there is a data gap in the Si IV spectral image (see Fig. 2, panel (c)) in the Y-direction from 370 to 387 pixels.

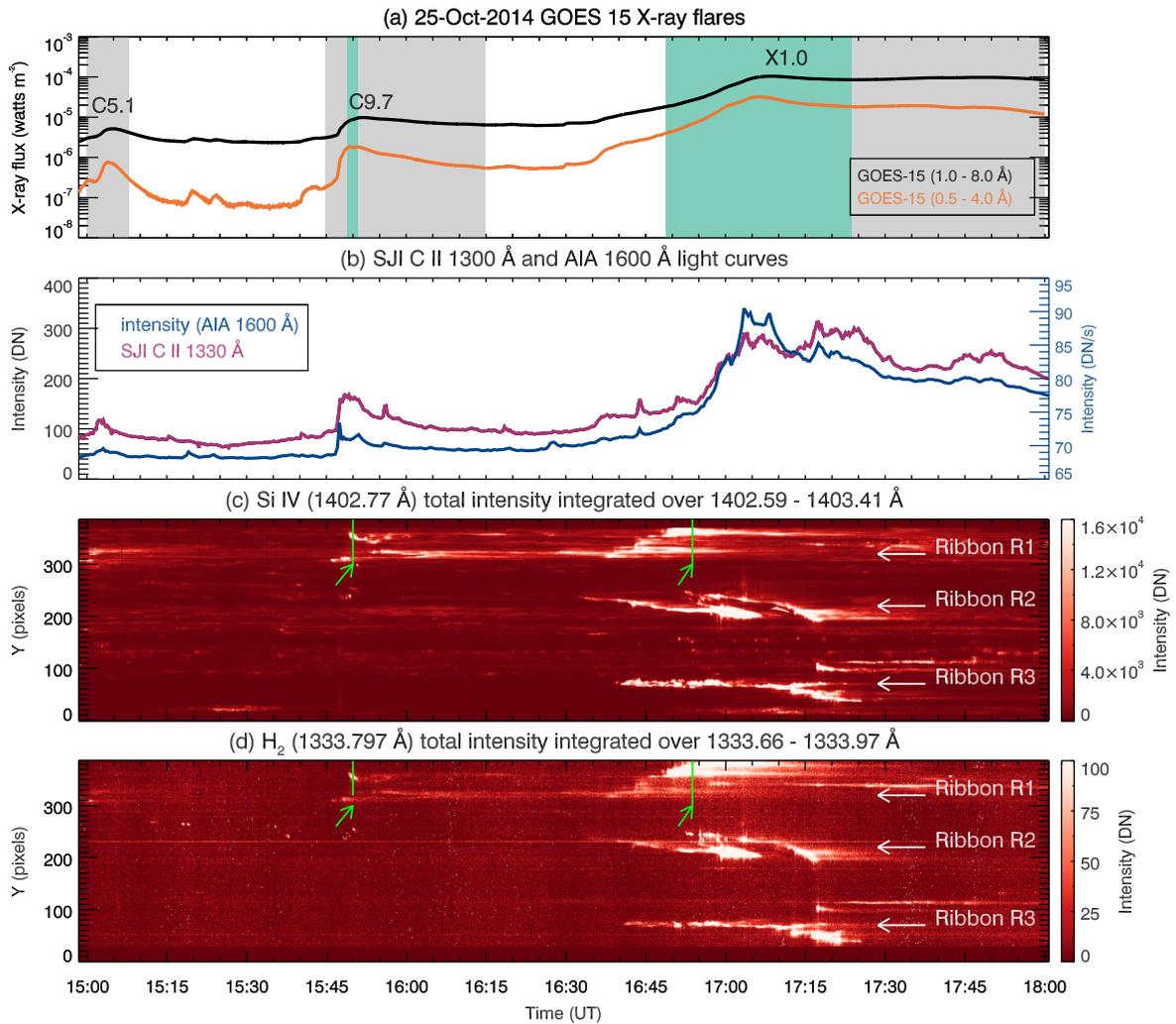

**Figure 2.** Panel (a): GOES X-ray fluxes for three flares, the same as is shown in panel (a) of Fig. 1. The grey shaded areas indicate the durations of each GOES flare, and the green shaded areas indicate the time slots where we were able to perform line fitting for $H_2$ emission during the flares. Panel (b): The light curves for the AIA 1600 Å (blue curve) and SJI 1330 Å (purple curve) channels were obtained by averaging the intensities over pixels covered by the FOV shown in Fig. 1, panels (d) and (g) respectively. Panels (c–d): The IRIS spectral images using the total intensity of the lines, obtained by summing over the wavelength ranges from 1402.59 to 1403.41 Å for the Si IV line and from 1333.66 to 1333.97 Å for the $H_2$ line. The emission from flare ribbons is seen as a white patchy structure and is shown by white arrows. The solid green lines shown by green arrows are the slit positions for which the spectra were obtained in Fig. 3.

**Table 2.** IRIS flare observation details

| IRIS | Spectrograph (SG) | Slit-Jaw-Imager (SJI) |
|---|---|---|
| Observation date | 25−Oct−2014 | |
| Observation ID | 3880106953 | |
| NOAA* active region | #12192 | |
| Start time (UT) | 14:58:28 | |
| End time (UT) | 18:00:47 | |
| No. of rasters | 1 | 606 (No. of images) |
| No. of slit positions | 2040 | – |
| Roll angle (slit) | +90° | – |
| Step cadence (sec) | 5.4 | – |
| Spatial resolution | 0.33″ (slit width) | 0.33″ |
| Field-Of-View (FOV) | 0.33″× 129″ | 121″× 129″ |
| Exposure time (sec) | 3.9 | 1.95 |
| Cadence | 5.4 sec (for each raster) | 16 sec (for C II, Mg II) |

**Notes -** *NOAA − National Oceanic and Atmospheric Administration

## 3 OVERVIEW OF THE THREE FLARES

In this section, we give an overview of the characteristics and evolution of each of the three flares in the series. The detailed spectroscopic analysis in the coming sections will focus primarily on the emission at the Ribbon 1 location during the X1.0 flare event.

### 3.1 Development of Flare 1: GOES C5.1

The C5.1 flare (see Figs. 1 and 2, panel (a)) started at 15:00 UT, peaked at 15:04 UT and ended at 15:08 UT (GOES 1.0−8.0 Å), with some pre-flare activity from 14:53 UT. In the IRIS UV light curve, the C II intensity increases slowly from 14:58:28 UT towards two peaks at 15:02:20 UT and 15:03 UT followed by a slow decay (see Fig. 2, panel (b)). The AIA 1600 Å channel behaves similarly, with a single peak at 15:03 UT. As would be expected for a chromospheric

**Table 3.** Details of H$_2$ emission lines observed by IRIS

| (Column 1) | (Column 2) | (Column 3) | (Column 4) | (Column 5) | (Column 6) | (Column 7) | (Column 8) |
| --- | --- | --- | --- | --- | --- | --- | --- |
| Exciting line $\lambda$ (Å) | Upper level ($v' - j'$) | Lower level ($v'' - j''$) | Branch ($\Delta J = \pm 1$) | H$_2$ $\lambda$ (Å) | Oscillator Strength | Transition probability A (s$^{-1}$) | Branching ratio |
| Si IV 1393.76 | 0–1 | 4–0 | R0 | 1333.475 | 0.1251 | 1.56316e+08 | 0.51 |
|  | 0–1 | 4–2 | P2 | 1338.565 | 0.04971 | 3.08291e+08 |  |
|  | 0–1 | 5–0 | R0 | 1393.719 | 0.1183 | 1.35378e+08 | 0.52 |
|  | 0–1 | 5–2 | P2 | 1398.954 | 0.04626 | 2.62639e+08 |  |
| Si IV 1402.77 | 0–2 | 4–1 | R1 | 1333.797 | 0.08355 | 1.87859e+08 | 0.68 |
|  | 0–2 | 2–3 | P3 | 1342.257 | 0.05322 | 2.75700e+08 |  |

**Notes** – Vibrational quantum number, $v'$ (upper level) and $v''$ (lower level), Rotational quantum number, $j'$ (upper level) and $j''$ (lower level). For P transition, $\Delta J = J' - J'' = -1$ and for R transition, $\Delta J = J' - J'' = +1$. The above parameters were taken from the molecular spectroscopy database, SESAM – http://sesam.obspm.fr/. The pairs of H$_2$ lines grouped together have the same upper level (see column 2).

signature, the peaks occur before the SXR peak time of the C5.1 flare.

Ribbon R1 develops from a chromospheric brightening between the two main sunspots (see Fig. 1, panel (d), X-axis = 415″ to 425″ and Y-axis = -315″ to -325″). Observed in the AIA 1600 Å channel this begins at 14:58:16 UT (∼2 min before the GOES start time of the C5.1 flare). The brightness here increased and the ribbon elongated to the northeast (NE) from this location between 15:03:04 UT and 15:06:40 UT when it disappeared. Some small-scale brightening was also observed in the penumbra of the trailing spot. The SJI C II images also show transition region brightenings at the same location (see Fig. 1, panel (g), X-axis = 415″ to 450″ and Y-axis = -295″ to -330″) with a similar ribbon appearance and development between 15:03:02 UT and 15:04:55 UT, and penumbral bright patches.

The spectrograph captured emission from ribbon R1 in TR lines, Si IV (see Fig. 2, panel (c)), O IV and cool chromospheric lines such as C I, and O I. Though emission from the low-temperature molecular hydrogen H$_2$ lines was visible (see Fig. 2, panel (d), between 15:00–15:04 UT, Y-pix = 300–330) it was too weak to fit the lines satisfactorily. Si IV emission at ribbon R1 during the C5.1 flare was not as bright as the emission seen at ribbon R1 during the C9.7 and X1.0 flares.

### 3.2 Development of Flare 2: GOES C9.7

The GOES C9.7 flare start, peak and end times are 15:44 UT, 15:52 UT and 16:15 UT (see Figs. 1 and 2, panel (a)) respectively. The pre-flare X-ray fluxes started to increase slowly from 15:40 UT until 15:44 UT and then rapidly between 15:47 UT and 15:52 UT. The GOES 0.5–4.0 Å channel shows two peaks at 15:49:20 UT and 15:51 UT, ∼1.5 minutes before the peak time in the 1.0–8.0 Å channel. The AIA 1600 Å and SJI C II light curves (see Fig. 2, panel (b)) increase slowly from 15:25 to 15:46:40 UT, then rapidly. The R1 region between the two sunspots brightened again at 15:38:16 UT in the AIA 1600 Å channel (see Fig. 1, panel (e)) and ribbon R1 started to grow in the NE and south-west (SW) directions, extending in length and width beyond what was observed in the C5.1 flare. The SJI C II light curve shows multiple small impulsive peaks; the first of these at 15:47:30 UT coincides with the single peak in the AIA 1600 Å light curve, and corresponds to a brightening at the penumbra of the trailing sunspot. Another C II brightening at 15:47:52 UT at R1 shows rapid changes in the intensity of the entire ribbon R1 until the end of the flare.

During the C9.7 flare, the H$_2$ emission at ribbon R1 could be fitted for 19 slit positions during the impulsive phase of the flare, from 15:49:16–15:50:52 UT (1 min 36 sec). Sample spectra of Si IV and five H$_2$ lines during this flare are shown in the top panel of Fig. 3 (see panels (a–f)).

### 3.3 Development of Flare 3: GOES X1.0

The GOES X1.0 flare start, peak and end times are 16:55 UT, 17:08 UT and 18:11 UT (see Figs. 1 and 2, panel (a)) respectively. The rising phase of the X1.0 flare was a bit longer compared to the other two flares. There is a preceding slow increase in the X-ray fluxes from 16:36 UT. The intensities in AIA 1600 Å and SJI C II channels show multiple impulsive peaks, starting at around 16:44 UT.

From ∼30 minutes before the start of the flare, there was bright emission at ribbon R1 in AIA 1600 Å images and faint emission (similar to background plasma) in the SJI C II channel. Ribbon R1 brightened up at 16:31:52 UT and started to grow southwards of the leading sunspot until 16:47:28 UT. Ribbon R2 started to appear between the two sunspots at 16:32:40 UT at X-axis = 420″ and Y-axis = -330″ and started to grow in the NE direction towards the northern part of the trailing sunspot at 16:38:16 UT. During the flare rise phase, ribbon R2 also moved eastward. Once both ribbons R1 and R2 reached maximum lengths, they started to increase rapidly in brightness at the GOES start time, 16:55 UT. At this time, a third small brightening in the northern part of the trailing sunspot started to extend to the north, becoming Ribbon R3. The width of the three flare ribbons increased with time. All three flares were very bright at 17:06:40 UT. Ribbons R1 and R3 started to grow again, reaching their maximum length by the flare end time. Ribbon R2 disappeared first at 17:53:28 UT, followed by ribbon R1 and R3.

Emission in Si IV and H$_2$ lines (see Fig. 2, panels (c–d)) was observed at ribbons R1, R2, and R3 and shown as a white patchy structure indicated by white arrows. The H$_2$ spectral lines at ribbon R1 were fitted during the impulsive rise of the flare from 16:48:07–16:55:04 UT. At R2, H$_2$ spectral lines were fitted from 17:03:23–17:06:41 UT, close to the peak time of the flare. H$_2$ spectral lines at R3 were observed from the start time of the flare until 17:21:48 UT (31 min 23 sec) and covered the peak time of the flare as well.

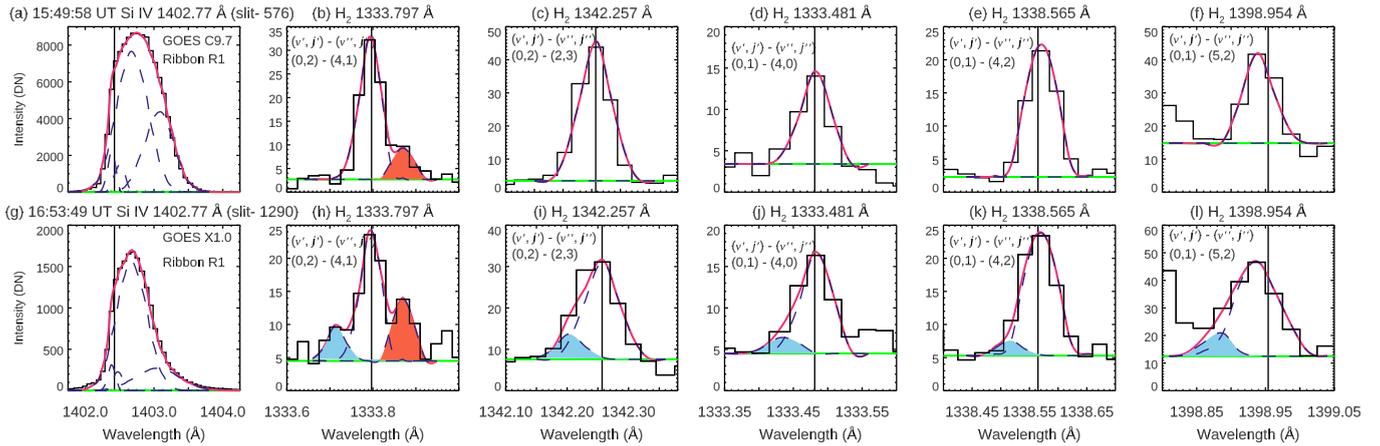

**Figure 3.** Various H$_2$ and Si IV 1402.77 Å spectral lines fitted during the impulsive rise phases of the C9.7 (top panel, a–f) and X1.0 (bottom panel, g–l) flares at ribbon R1. These spectra were obtained for the slit number 576 (1290) for C9.7 (X1.0) flare and averaged over the pixels along the slit 350−356 (365−371) for C9.7 (X1.0). These slit locations are shown as solid green lines and indicated by green arrows in Fig. 2, panels (c–d). The spectra are shown as black histograms. The Gaussian components are shown with blue dashed lines with a shaded region, whereas the total fit is shown by solid pink lines. The green lines indicate the background emission. The blue-wing (red-wing) component of H$_2$ is highlighted by the sky-blue (red) colour. The black solid vertical line indicates the theoretical wavelength of the spectral lines obtained from Sandlin et al. (1986). The upper (lower) levels of the transitions for each H$_2$ are indicated where vibrational quantum number, $v'$ (upper level) and $v''$ (lower level), rotational quantum number, $j'$ (upper level) and $j''$ (lower level).

## 4 SPECTRAL ANALYSIS OF THE X1.0 FLARE

In this section, we discuss the behaviour of the spectral lines observed at the ribbon R1, primarily during the X1.0 flare, and the properties of the emitting material that can be deduced from the observations.

### 4.1 Line profiles of Si IV and five H$_2$ lines

Figure 3 shows an example of spectral profiles for one Si IV line at 1402.77 Å (panels (a) and (g)) and five H$_2$ lines (panels b–f and h–l) during the impulsive phases of the C9.7 (top panel) and X1.0 (bottom panel) flares. The spectra shown are at one slit position (i.e. time) at ribbon R1 during each flare i.e. at slit number 576 (1290) and averaged over the pixels 350−356 (365−371) along the slit for the C9.7 (X1.0) flare. These slit positions are shown as solid green lines (and also indicated by green arrows) in Fig. 2, panels (c–d). In Fig. 3, the vertical black solid lines indicate the theoretical wavelengths of the spectral lines obtained from Sandlin et al. (1986). The total fit is shown by solid pink lines, and the green lines indicate the background emission. In the X1.0 flare, the H$_2$ emission was fitted with Gaussians for a few minutes only, between 16:48:07−16:55:04 UT in R1, between 17:03:23−17:06:41 UT in R2, and between 16:50:25−16:59:27 UT in R3.

The Si IV profiles (Fig. 3, panels (a) and (g)) are quite broad and need multiple Gaussians to fit them. Similar broad Si IV profiles were also reported at ribbons in some other solar flares, and multi-Gaussian fits were applied (Brannon et al. 2015; Lörinčík et al. 2022a,b). The H$_2$ lines in panels (b) through (f) are grouped by the shared upper level of the transition. In panels (b) and (c), the upper level is $(v', j') = (0, 2)$, and in panels (d) to (f), it is $(v', j') = (0, 1)$. Panels (h) through (l) are corresponding examples from the X1.0 flare.

All five H$_2$ spectral lines during the flares have a "stationary" component (Gaussian fit shown with blue dashed lines in Fig. 3). The fitted wavelength is equal to the expected rest wavelength (obtained from Sandlin et al. (1986) and shown as black vertical lines), within errors. However, the H$_2$ line at 1333.797 Å, $(v', j') = (0, 2)$ also shows a component in the red wing (highlighted as a red Gaussian in Fig. 3, panel (b)) in addition to the stationary component. This is also seen in the X1.0 flare (panel (h)). We cannot identify this red-wing component in any line lists or databases. We checked whether there is a possible blend of S I at 1333.80 Å with the H$_2$ at 1333.797 Å by following the procedure given in Appendix B of Mulay & Fletcher (2021) and we found that the contribution from S I is not present in this line. The red wing component in H$_2$ 1333.797 Å could also be red shifted emission in the $(0, 2) − (4, 1)$ transition, with a line-of-sight speed of around 20 km s$^{-1}$. A moving component in the blue wing was present for all five H$_2$ profiles in addition to the stationary component, at a speed between 10−20 km s$^{-1}$.

The temporal evolution of intensities at ribbon R1 for one Si IV and four H$_2$ spectral lines during the X1.0 flare, during the time interval for which Gaussian fits could be reliably carried out, is shown in Fig. 4. The purple curves are the H$_2$ intensities obtained by summing the intensities over the wavelength ranges considering the extent of the lines (note that these wavelength ranges include the red and blue wing components observed). The black curves indicate intensities obtained from the Gaussian fits of the stationary component of H$_2$. The intensity variation in the stationary components in four of the H$_2$ lines are very similar, but differ in the blue- and red-wing Gaussian components. In the case of the H$_2$ lines shown in panels (b) and (d), the H$_2$ line at 1333.797 Å has more of a red wing component whereas the other H$_2$ line at 1342.257 Å has more of a blue wing component, even though the same Si IV emission at 1402.77 Å is responsible for exciting both H$_2$ lines.

### 4.2 Doppler and nonthermal velocities of four H$_2$ lines

We measured the Doppler and nonthermal velocities for four of the five H$_2$ lines at ribbon R1 during X1.0 flare, and the results are shown in Figs. 5 and 6. The black curves indicate velocities for the stationary components of the H$_2$ lines. The velocities of the blue (red) wing components are shown as blue (red) curves. The Doppler velocities of the stationary components of these four H$_2$ lines indicate negligible bulk flows along the line of sight. Small red-shifts of ∼10–15 km s$^{-1}$ correspond to downflows, which were observed for the red wing component of the H$_2$ line at 1333.797 Å. For other H$_2$ lines, small red and blue shifts of ∼5–10 km s$^{-1}$ were observed.

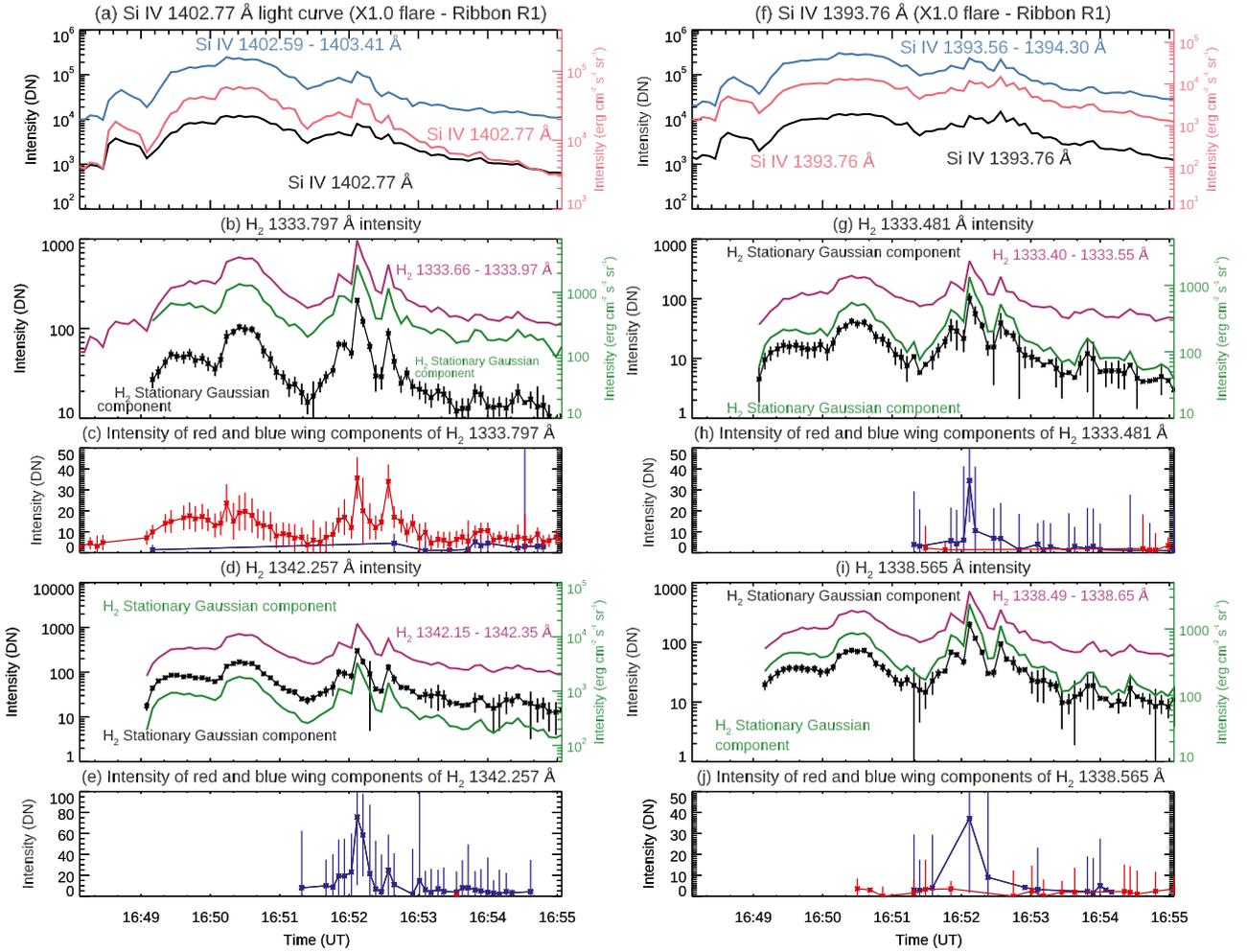

**Figure 4.** The behavior of Si IV and H$_2$ at ribbon R1 location during the X1.0 flare. Panels (a) and (f): Temporal evolution of Si IV 1402.77 and 1393.76 Å lines at a single wavelength value is shown as the black and pink curves. The blue curves indicate the total intensities obtained by summing over the wavelength ranges (around Si IV 1402.77 and 1393.76 Å lines) mentioned in the plots. Panels (b–e): Temporal evolution of H$_2$ lines 1333.797 Å, and 1342.257 Å which strongly reflect the absorption of photons from Si IV 1402.77 Å lines. Panels (g–j): Temporal evolution of H$_2$ lines 1333.481 Å, and 1338.565 Å which are produced due to absorption of photons from the Si IV 1393.76 Å lines. The black (DN units), pink and green (erg cm$^{-2}$ s$^{-1}$ sr$^{-1}$ units) curves are the peak intensities obtained by fitting the stationary component with a single Gaussian. The purple curves in all these panels are the total intensities obtained by integrating over the wavelength ranges mentioned in the plot. The red and blue curves (in panels c, e, h and j) indicate the temporal evolution of red and blue wing components observed in the H$_2$ line profiles. The intensities were obtained by fitting these components with a single Gaussian same as shown in Fig. 2, panels (h–l).

For the measurement of nonthermal velocities, $W_{nth}$, we used $W_{nth} = \sqrt{W_{1/e}^2 - W_{th}^2 - W_{instr}^2}$ where $W_{1/e}$ is the measured 1/e spectral line width. We fit the H$_2$ lines with single and/or multiple Gaussian components. We obtained the full-width half maximum (FWHM) of the line and converted it into 1/e spectral line width using FWHM = $2\sqrt{\ln 2}$ $W_{1/e}$ = 1.665 $W_{1/e}$ (Peter 2010). $W_{th}$ is the thermal line width of the H$_2$ line and is defined as $\sqrt{2k_B T/M}$, where $k_B$ is the Boltzmann constant, $1.380649 \times 10^{-23}$ J/K. We assume the temperature $T$ of formation of H$_2$ emission to be 4200 K (Innes 2008) and mass $M$ of the H$_2$ molecule is $3.34 \times 10^{-27}$ kg. We obtained $W_{th}$ = 5.88 km s$^{-1}$. $W_{instr}$ is the 1/e IRIS instrumental width, 3.51 km s$^{-1}$ measured using the formula given in the iris_nonthermalwidth.pro[6] routine (where IRIS instrumental FWHM is 26 mÅ for the FUV channel, De Pontieu et al. (2014). Fig. 6 shows nonthermal velocities for four H$_2$ lines. The black curves indicate nonthermal velocities for the stationary components, and the nonthermal velocities of the blue (red) wing component of H$_2$ lines are shown as blue (red) curves. As in Mulay & Fletcher (2021), significant nonthermal broadening was also found. The nonthermal velocities range from ∼10–20 km s$^{-1}$, and presumably represent the properties of the deep atmosphere where the H$_2$ molecules are found, and the line fluoresces.

---

[6] https://www.heliodocs.com/xdoc/xdoc_print.php?file=$SSW/packages/mosic/other_libs/iris_lib/iris_nonthermalwidth.pro

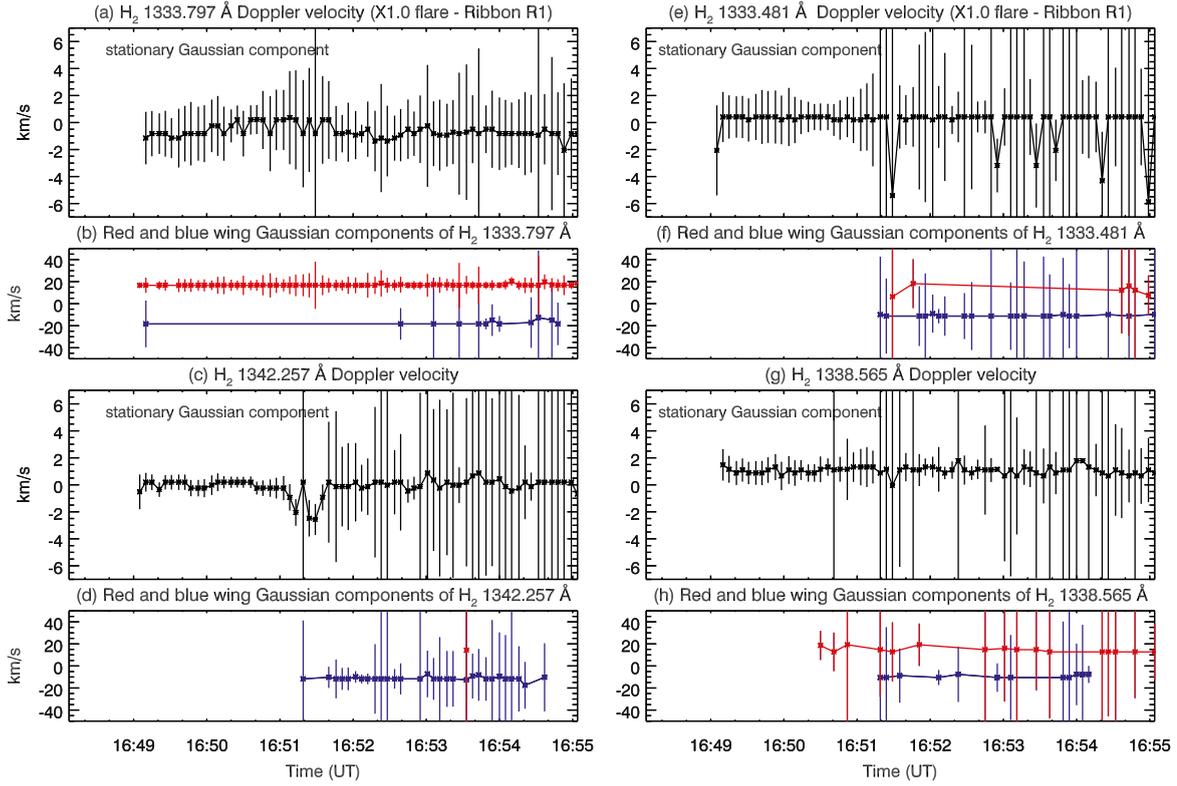

**Figure 5.** Doppler velocities of H$_2$ lines at ribbon R1 location during the X1.0 flare. The black curves are the Doppler velocities for a stationary component of H$_2$ lines. Doppler velocities for the red-wing (blue-wing) Gaussian component are shown as a red (blue) curve.

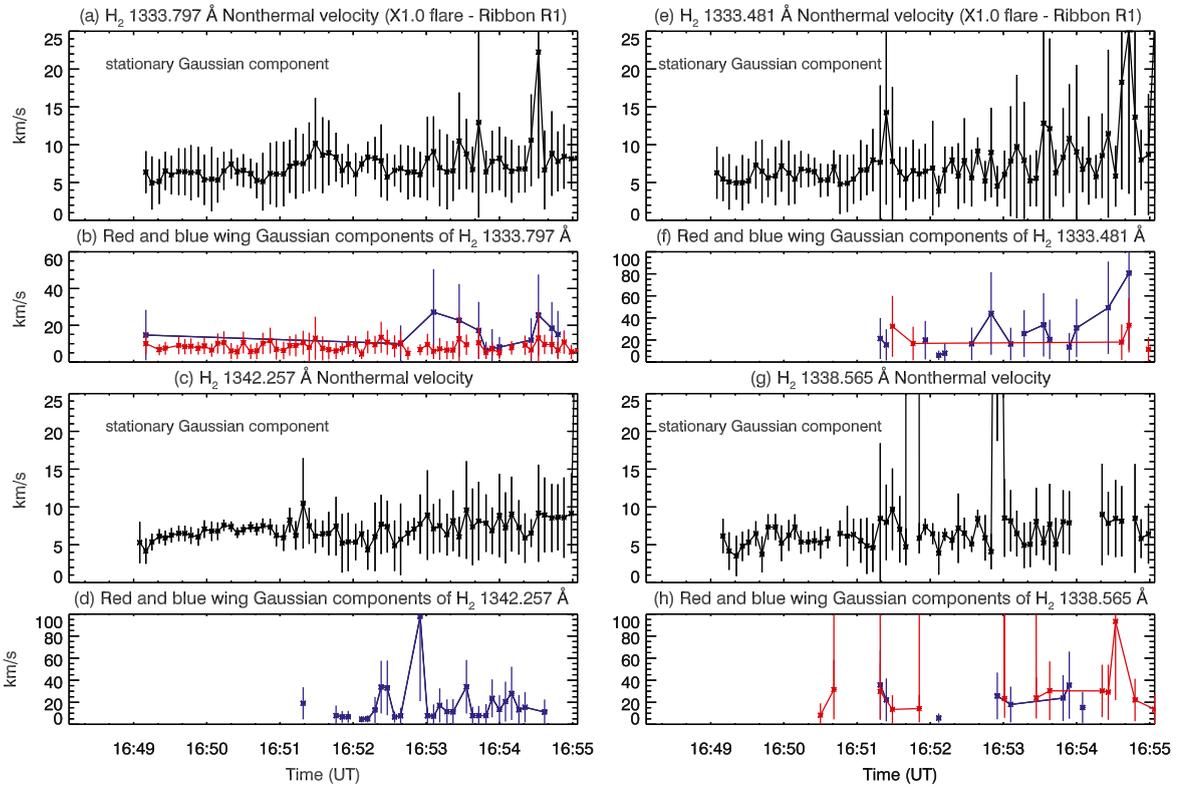

**Figure 6.** Nonthermal velocities of H$_2$ lines at ribbon R1 location during the X1.0 flare. The black curves are the nonthermal velocities for a stationary component of H$_2$ lines. Nonthermal velocities for the red-wing (blue wing) component are shown as a red (blue) curve.

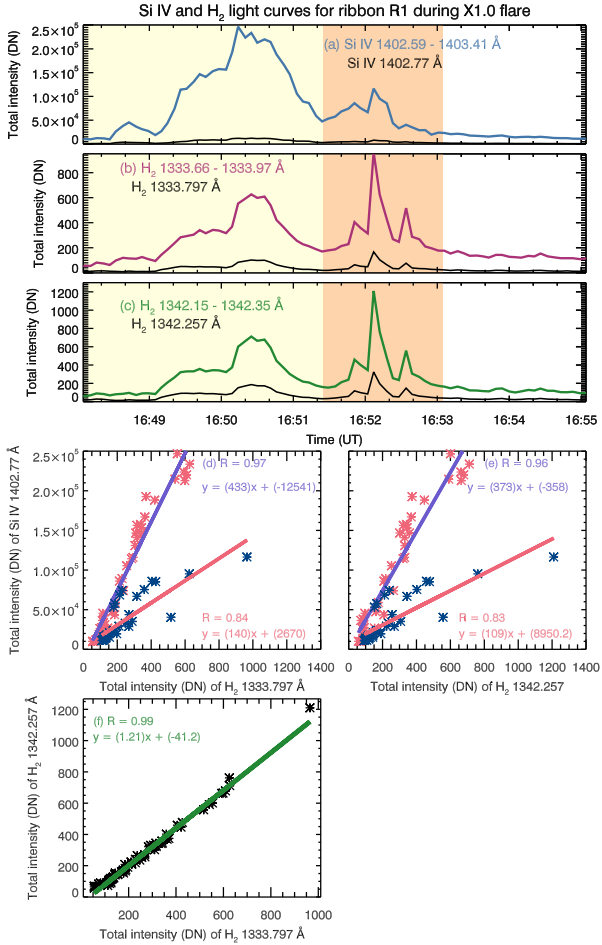

**Figure 7.** Temporal evolution of Si IV and H$_2$ emission at ribbon R1 during the X1.0 flare. Panel (a): Temporal evolution of Si IV (blue curve) (the intensity summed over 1402.59 and 1403.41 Å) and at a single wavelength value of Si IV 1402.77 Å (black curve). Panel (b–c): The temporal evolution of two H$_2$ lines (originating from the same upper level, $(v', j') = (0, 2)$) at a single wavelength value (black curve) and summed over a range of wavelengths (purple and green curves). The intensity variation was observed for the two time slots highlighted with yellow and orange backgrounds. Panels (d–e): Scatter plots for ribbon R1 obtained for the emission seen in Si IV (between 1402.59 and 1403.41 Å), H$_2$ (between 1333.66 and 1333.97 Å) and H$_2$ (between 1342.15 and 1342.35 Å) lines. The intensities are displayed with pink, and blue star symbols. The blue (pink) solid line indicates the linear fit to the data shown in the yellow (orange) background. Panel (f): Scatter plot for ribbon R1 obtained for the emission seen in two H$_2$ lines. The intensities are displayed with black star symbols, and the linear fit to the data is shown by a solid green line. The equations for the fitted lines along with fit parameters are given and the Pearson correlation coefficients are displayed as 'R'.

### 4.3 Correlation between H$_2$ and Si IV emission

Different H$_2$ lines are excited by different wavelengths of UV emission. The lines at 1333.797 Å and 1342.257 Å share upper level $(v', j') = (0, 2)$ and are fluoresced by Si IV 1402.77 Å. In Mulay & Fletcher (2021) we found a good correlation between the intensities in H$_2$ 1333.797 Å and Si IV 1402.77 Å which we investigate further here. Fig. 7, panels (a–c), shows the temporal evolution of Si IV and these two H$_2$ lines at ribbon R1 (time interval − 16:48:07–16:55:04 UT) during the X1.0 flare. Panels (d–f) show the scatter plots for Si IV and H$_2$ emission. We observed that there were two different correlations present between the line intensities in two-time slots that are shown as yellow- and orange-shaded regions. We obtained two correlation coefficients by fitting two lines (pink and blue lines). In panels (d) and (e), the intensities shown by pink stars are obtained from the first time slot (shown in the yellow-shaded region) and the blue line shows the fit. The intensities shown by the blue stars are obtained from the second time slot (shown in the orange-shaded region) and the pink line shows the fit. The intensities of the fluoresced and exciting radiation are strongly correlated, though in R1 the slope of the correlation is different at different times. This implies some complexity in the nature of the UV excitation. We also found a strong correlation between two H$_2$ lines at 1333.797 Å and 1342.257 Å which is shown in panel (f).

We investigated the correlation between the exciting Si IV 1393.76 Å line and its fluoresced H$_2$ emission at 1333.481, 1338.565, and 1398.954 Å which share the same upper level $(v', j') = (0, 1)$. The results are shown in Fig. A1 in the Appendix. We found that there is a strong spatial and temporal correlation between Si IV and these three H$_2$ emission lines at ribbon R1, as well as a strong correlation between the three H$_2$ lines. A similar analysis was carried out for ribbon R2 and R3 and the results are summarized in Table 4.

### 4.4 Branching ratios

Based on selection rules, for a given upper energy level there are many different possible downward transitions to lower energy levels. The intensity of a spectral line depends on the probability associated with the transition, e.g. more photons will be produced when the transition probability (A-value) is high. The branching ratio facilitates understanding of the relationship between intensities of two spectral lines originating from the same upper level. It is expressed as the ratio of the transition probability of a spectral line and the total probability of all the downward transitions from the same upper level. In the optically thin case, the intensity ratio of two spectral lines originating from the same upper level is equal to the ratios of their transition probabilities (A-values) (Jordan 1967; Del Zanna & Mason 2018). Observed ratios can then be compared with predicted branching ratios to assess plasma opacity.

We investigated how the measured intensities of four H$_2$ lines are related to their branching ratios (see Table 3, columns 7 and 8) and the plasma condition under which the H$_2$ emission is produced (Jordan et al. 1978; Jaeggli et al. 2018). Table 3 gives the wavelengths and contribution fractions to the upper-level population for the strongest pumped transitions to the upper level for every line listed in the table.

Figure 8 shows the peak intensities of the stationary components of the four H$_2$ lines and their intensity ratios. The H$_2$ lines indicated in pink have large A-values (resulting in high intensity) compared to the other H$_2$ lines shown in violet, where the A-values are smaller. We plotted the ratio of the A-values of the H$_2$ lines as the black horizontal lines. Panels (b) and (d) show that the ratio of the intensities of H$_2$ lines (blue curves) is equal to the ratio of their transition probabilities, A-values (black horizontal lines). These results confirm that most of the time the measured branching ratios agree within errors with the predicted values, meaning that the H$_2$ lines are produced under optically thin plasma conditions.

## 5 THE DEPTH OF THE H$_2$-EMITTING LAYER

In earlier sections, we showed evidence of the strong spatial and temporal correlation between Si IV and H$_2$ emission. These results indicate that the downward directed photons from two Si IV lines produce multiple H$_2$ lines through the fluorescence process. This implies that

**Table 4.** The spatial and temporal correlation between Si IV and $H_2$ emission at three ribbon locations R1, R2 and R3 during X1.0 flare

| (Column 1) | (Column 2) | (Column 3) | (Column 4) | (Column 5) | (Column 6) | (Column 7) | (Column 8) | (Column 9) |
|---|---|---|---|---|---|---|---|---|
| Exciting line | $H_2$ | Pearson correlation coefficient (R) | | | $H_2$ lines | Pearson correlation coefficient (R) | | |
| $\lambda$ (Å) | $\lambda$ (Å) | Ribbon R1 | Ribbon R2 | Ribbon R3 | $\lambda$ (Å) | Ribbon R1 | Ribbon R2 | Ribbon R3 |
| Si IV 1402.77 | 1333.797 | 0.97 (0.84) | 0.91 | 0.88 | 1333.797–1342.257 | 0.99 | 0.95 | 0.97 |
| Si IV 1402.77 | 1342.257 | 0.96 (0.83) | 0.97 | 0.88 | | | | |
| Si IV 1393.76 | 1333.481 | 0.96 (0.90) | 0.93 | 0.83 | 1333.481–1338.565 | 0.99 | 0.98 | 0.95 |
| Si IV 1393.76 | 1338.565 | 0.95 (0.89) | 0.95 | 0.81 | 1338.565–1398.954 | 0.96 | 0.98 | 0.85 |
| Si IV 1393.76 | 1398.954 | 0.97 (0.93) | 0.94 | 0.78 | 1398.954–1333.481 | 0.98 | 0.98 | 0.89 |

**Notes** – In column 3, the numbers in the parentheses indicate the correlation coefficients obtained for the time slots shown as orange colour background in Fig. 7.

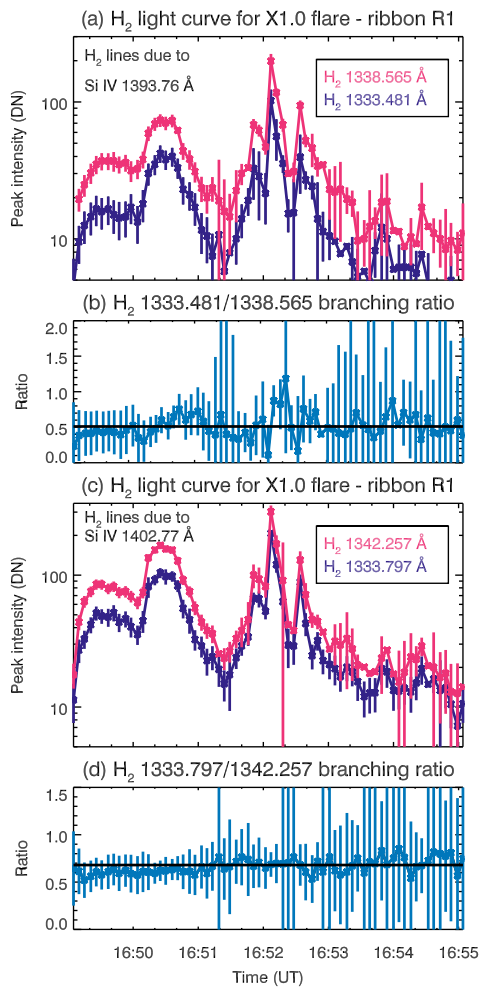

**Figure 8.** Panel (a) and (c): $H_2$ peak intensities for the stationary components of the 1333.481, 1333.565, 1333.797 and 1342.257 Å lines at ribbon R1 during the X1.0 flare. Panels (b) and (d): The intensity ratio of $H_2$ lines is shown as blue curves and the ratio of the transition probabilities, A-values (see Table 3, column 8) of this $H_2$ lines (i.e. branching ratio) is plotted as black horizontal lines. The vertical lines in all panels indicate the error bars on the ratio of the $H_2$ peak intensities.

the Si IV emission, which becomes very intense during flares, could heat the lower atmosphere through radiative back-warming (Machado et al. 1989; Metcalf et al. 1990; Allred et al. 2005). It also allows us to apply a geometric model developed by Isobe et al. (2007), to estimate the depth of the $H_2$ emitting layer relative to the source of the exciter radiation.

The three ribbons observed in the spectral images (see Fig. 2, panels (c)) during the X1.0 flare show sharp edges in the Si IV spectral image. The resultant fluorescence $H_2$ emission (see Fig. 2, panels (d)) was found to be bright where Si IV emission is strong. In addition to this there is faint and diffuse $H_2$ emission around the three flare ribbons. This is consistent with radiative back-illumination or back-warming, which we would expect to have a characteristic "core-halo" emission structure (Neidig et al. 1993; Hudson et al. 2006).

A similar investigation of the radiative back-warming effect as an explanation of the strong flare optical continuum was carried out by Isobe et al. (2007) in a flare source showing just such a "core-halo" structure. Using the intensity profile of a flare ribbon projected against a sunspot umbra, they estimated the depth of the diffuse "halo" flare emission to be 100 km or less from the height of the source of radiative back-warming. We followed this method to study the location and depth of diffuse $H_2$ emission from the height of the Si IV emission.

We focused on ribbon R1 during the X1.0 flare. Fig. 9 shows the spectral images for ribbon R1 in $H_2$ and Si IV lines (same as that shown in Fig. 2, panels (c) and (d) but zoomed in on ribbon R1). The $H_2$ emission is over plotted in both panels of Fig. 9 as black contours. We selected Y-pixels from 300 to 370 to achieve sufficient signal in the selected $H_2$ line we binned over three regions corresponding to slit numbers – 1250–1300 (16:50:14–16:54:43 UT), 1300–1350 (16:54:43–16:59:11 UT), and 1350–1400 (16:59:11–17:03:39 UT) shown as three green boxes in Fig. 9, along the ribbon R1 where we observed the sharp boundaries of the ribbon in the Si IV spectral image (panel (b)) and the diffuse region around the ribbon in the $H_2$ spectral image (panel (a)). Fig. 10 shows time-averaged intensities for $H_2$ and Si IV lines that were obtained at slit positions 1250–1300 (top panel), 1300–1350 (middle panel), and 1352–1400 (bottom panel) for Y-pixel numbers from 300 to 370. The orange-shaded areas show the intensity profiles observed for $H_2$ and Si IV emission where the radiative back-illumination effect was investigated. These areas indicate regions where the $H_2$ sources are noticeably more extended along the slit than the associated Si IV sources, consistent with the diffuse $H_2$ emission being caused by radiative back-illumination.

We followed the simple geometric model given by Isobe et al. (2007) (see section 3.1 in that paper) to estimate the depth of the diffuse $H_2$ emission using the time-averaged intensity profiles of $H_2$ emission. This model expresses the spatial variation of the intensity of the emission caused by back-illumination (in our case the fluoresced H$_2$ emission) in terms of the irradiation from the primary source (in our case the Si IV emission), the displacement $h$ between the primary and back-illumination source (see Fig. 6 in Isobe et al. (2007), who use the term "height"), and an additional coefficient. We selected two time-averaged H$_2$ intensity profiles from the middle and bottom panels of Fig. 10 within the X-axis range from 350 (350) to 362 (361) for slit numbers 1300–1350 (1350–1400) and plotted them in Fig. 11. We used Equation 2 from Isobe et al. (2007) to perform least-squares fitting of the intensity profiles and obtained the depth of H$_2$ emission to be $h$ = 1871±157 km (for slits 1300–1350, 16:54:43–16:59:11 UT) and 1207±112 km (for slits 1350–1400, 16:59:11–17:03:39 UT)) from the source of the Si IV emission placing it deep in the chromosphere, as would be expected for a molecular line. We note that the model assumes a point-like source of irradiation, whereas the Si IV emission is extended, however, it is substantially more localised than the H$_2$ so the model is a reasonable approximation. A diffuse EUV source will broaden the backwarmed area, as would finite thicknesses of either source and so the estimates here are upper limits on the actual displacement between the EUV and H$_2$ emission layers.

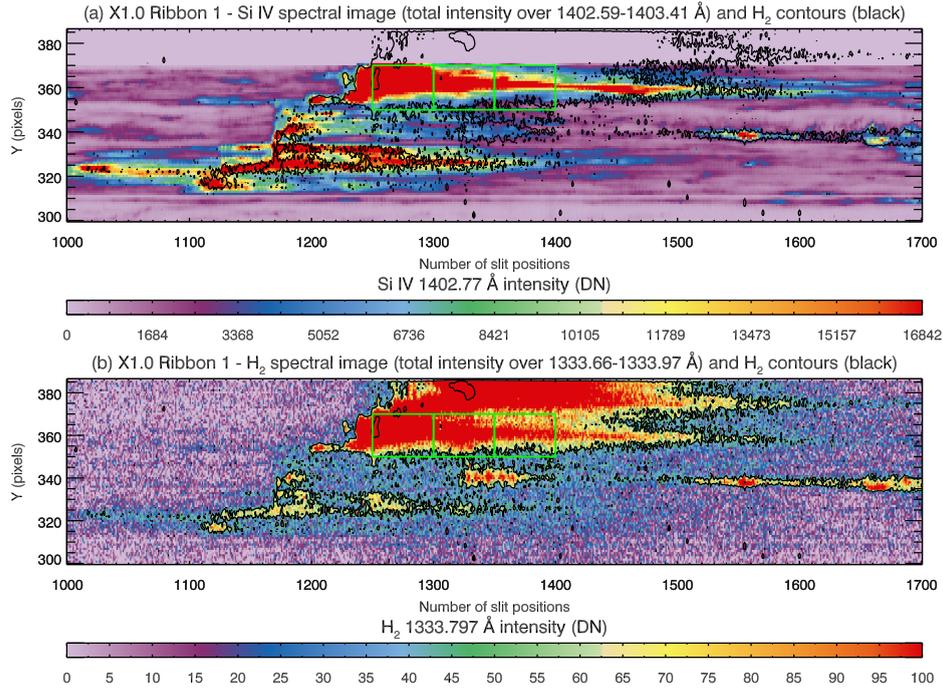

**Figure 9.** Spectral images for Si IV (panel a) and H$_2$ (panel b) emission. The images were created using the total intensities summing over the 1402.59 and 1403.41 Å for Si IV and 1333.66 and 1333.97 Å for H$_2$. The black contours in both panels indicate the contours for the H$_2$ emission. The intensities along the slit positions (box 1: 1250−1300 (16:50:14−16:54:43 UT), box 2: 1300−1350 (16:54:43−16:59:11 UT), and box 3: 1350−1400 (16:59:11−17:03:39 UT)) enclosed in three green boxes were chosen for the analysis. The H$_2$ and Si IV emission at these slit numbers was used to get the averaged intensities shown in Fig. 10. Note that there is a data gap in Si IV image between Y-pix = 370−387 for all slit positions.

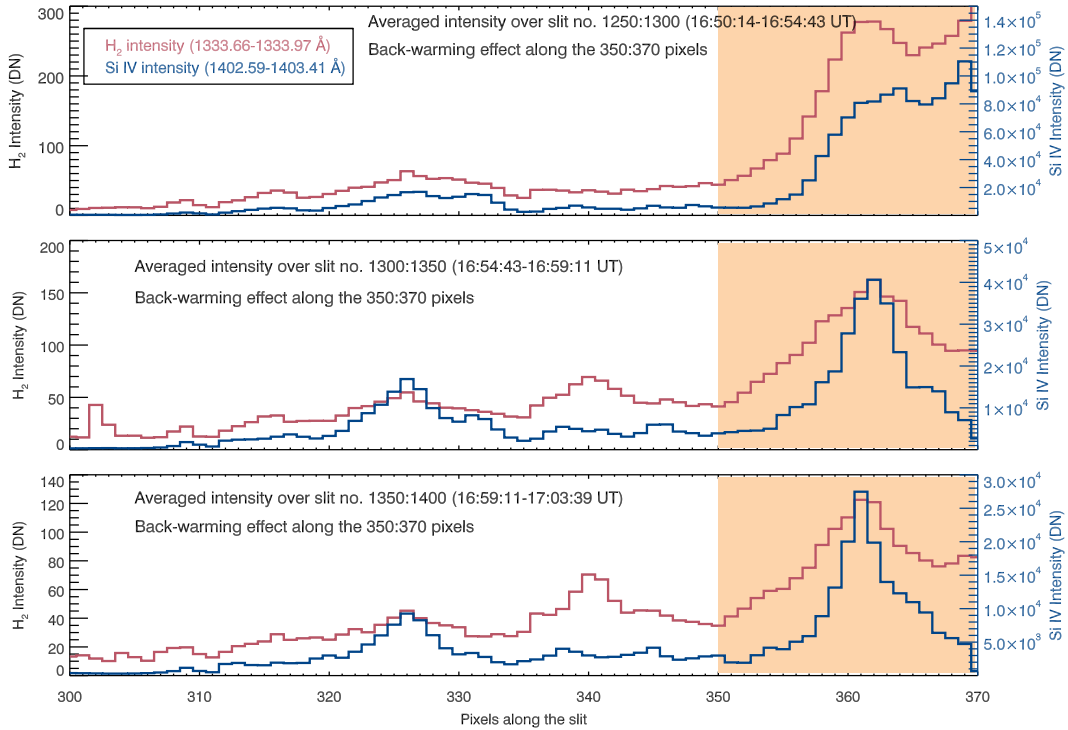

**Figure 10.** The time averaged intensities for H$_2$ and Si IV lines obtained at slit positions, 1250−1300 (top panel), 1300−1350 (middle panel), and 1352−1400 (bottom panel) for the pixel numbers from 300 to 370 (these are Y-pixels shown in Fig. 9). The orange-shaded areas indicate the intensity variation observed for H$_2$ and Si IV emission where the back-illumination effect was investigated.

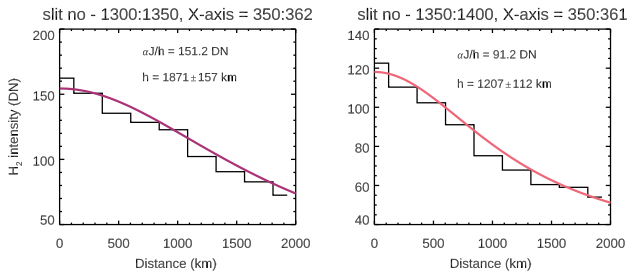

**Figure 11.** Histogram indicates the time-averaged H$_2$ intensities for slit positions, 1300–1350 (left panel, indicates time – 16:54:43–16:59:11 UT), and 1350–1400 (right panel, indicates time – 16:59:11–17:03:39 UT) as a function of distance, i.e. Y-pixels positions shown in Fig. 9 (that are same as the range of X-axis values – 350:362 (for the left panel), 350:361 (for the right panel) shown in Fig. 10). The least-square fitting was performed for the time-averaged H$_2$ intensities using the python *curve_fit* routine and fitted curves are shown in purple and pink colours. Here, we used 1 IRIS pixel = 0.33″ = 241.76 km to convert the Y-axis pixel locations shown in Fig. 9 to distance in km. The $h$ is the depth of the H$_2$ emission layer (with respect to the source of Si IV emission) excited by back-illumination, $J$ is the energy of irradiation per unit length and per radian and $\alpha$ is a coefficient (see equation 2 in Isobe et al. (2007)).

## 6 SUMMARY AND DISCUSSION

In this paper, we studied the temporal variation and spatial behaviour of cool molecular H$_2$ emission during C9.7 and X1.0 flares at a cadence of ~5.4 sec. Five H$_2$ lines were observed during the impulsive phases of the flares and studied thoroughly using plasma diagnostic techniques. The following plasma properties were derived from our analysis:

- The cool H$_2$ emission was observed at a ribbon in the impulsive rise phase of the GOES C9.7 X-ray flare. Strong H$_2$ emission was observed at three ribbons before the GOES X1.0 X-ray flare started and H$_2$ emission continued in the impulsive phase and beyond the peak time of the flare (see Fig. 2 and Table 1).

- In addition to a stationary component in five H$_2$ spectral profiles, we observed the presence of red wing and blue components indicating downflows and upflows respectively. Each component was fitted independently with a single Gaussian (see Fig. 3).

- We observed a strong spatial and temporal correlation between H$_2$ and Si IV emission indicating that two Si IV lines are responsible for five H$_2$ spectral lines (see Table 4).

- We observed a strong spatial and temporal correlation between two H$_2$ lines that have the same upper level and that are fluoresced by the same Si IV emission (see Table 4).

- The intensities of two H$_2$ lines with the same upper level are found to be consistent with their branching ratio (i.e. the ratio of their intensities was equal to the ratio of their transition probabilities, A-values, see Fig. 8). This indicates that the H$_2$ emission was formed under optically thin plasma conditions (see Table 3).

- The stationary components of the H$_2$ spectral profiles show negligible flows along the line-of-sight. Small redshifts of ~10–15 km s$^{-1}$ were measured for the red-wing component of the H$_2$ line at 1333.797 Å. For other H$_2$ lines, small red and blue shifts of ~5–10 km s$^{-1}$ were observed (see Fig. 5).

- The nonthermal velocities of H$_2$ range from ~5–15 km s$^{-1}$, and presumably represent the properties of the deep atmosphere during the flare where the H$_2$ molecules are found (see Fig. 6).

- Si IV spectral images showed relatively sharp ribbon edges during the X1.0 flare, whereas the H$_2$ emission shows faint and diffuse emission around the three flare ribbons (see Fig. 2, panels (d)) in addition to bright sources where Si IV emission is strong. These diffuse H$_2$ emission patches are consistent with radiative back-illumination (Isobe et al. 2007) of the lower layers by the Si IV emission (see Figs. 9 and 10).

- We estimated the displacement, $h$ of the diffuse H$_2$ emission with respect to the Si IV emission source using the time-averaged intensity profiles of H$_2$ and found $h$ to be 1871±157 km (for slits 1300-1350 that correspond to 16:54:43–16:59:11 UT), and 1207±112 km (for slits 1350–1400 that correspond to 16:59:11–17:03:39 UT) (see Fig. 11).

We believe that further study of the H$_2$ fluorescent emission lines, and derived plasma parameters, will be useful for defining models for the chromosphere and TMR during flares. The main limitation to our use of the fluorescent lines is their relatively low signal-to-noise ratio, which limits their image definition and time resolution, but we note the great significance of these layers in defining flare energy, momentum transfer, and magnetic fields.

## ORCID ID'S

Sargam M. Mulay  https://orcid.org/0000-0002-9242-2643
Lyndsay Fletcher  https://orcid.org/0000-0001-9315-7899
Hugh Hudson  https://orcid.org/0000-0001-5685-1283
Nicolas Labrosse  https://orcid.org/0000-0002-4638-157X

## ACKNOWLEDGEMENTS

SMM, LF and NL acknowledge support from UK Research and Innovation's Science and Technology Facilities Council under grant award number ST/T000422/1 and ST/X000990/1. The authors would like to thank Dr. Peter Young (NASA Goddard Space Flight Center, USA), Dr. Sarah Jaeggli (National Solar Observatory, USA), Dr. Alexander

MacKinnon (University of Glasgow, UK), Dr. Graham Kerr (The Catholic University of America, NASA Goddard Space Flight Center, Greenbelt MD), Dr. Christopher Osborne, Dr. Daniel Clarkson and Dr. Aaron Peat (University of Glasgow, UK), for discussion and valuable comments. SMM would like to thank Mr. Francesco Azzollini (University of Glasgow, UK) for helping her with Python programming and curve fitting routine. AIA data are courtesy of SDO (NASA) and the AIA consortium. NOAA Solar Region Summary data supplied courtesy of SolarMonitor.org. IRIS is a NASA small explorer mission developed and operated by LMSAL with mission operations executed at NASA Ames Research Center and major contributions to downlink communications funded by ESA and the Norwegian Space Centre. We acknowledge the RHESSI team for their support and open access to the data.

## DATA AVAILABILITY

In this paper, we used the Interactive Data Language (IDL) and SolarSoftWare (SSW; Freeland & Handy 1998) packages to analyze AIA, IRIS, RHESSI, and GOES data. All the figures within this paper were produced using IDL color-blind-friendly color tables (see Wright 2017). In addition, we used Python programming and packages for curve fitting. The AIA data is available at http://jsoc.stanford.edu/ and the data were analyzed using routines available at https://www.lmsal.com/sdodocs/doc/dcur/SDOD0060.zip/zip/entry/. IRIS has an open data policy. The IRIS data is available at https://iris.lmsal.com/data.html and the data[7] analysis was performed using the routines available at https://iris.lmsal.com/itn26/. The GOES data and analysis routines are available at https://hesperia.gsfc.nasa.gov/~kim/goes_software/goes.html. The RHESSI data and analysis routines are available at https://hesperia.gsfc.nasa.gov/rhessi3/data-access/rhessi-data/rhessi-data/index.html

## APPENDIX A: SOME EXTRA MATERIAL

This paper has been typeset from a T<sub>E</sub>X/L<sup>A</sup>T<sub>E</sub>X file prepared by the author.

---

[7] https://www.lmsal.com/hek/hcr?cmd=view-event&event-id=ivo%3A%2F%2Fsot.lmsal.com%2FVOEvent%23VOEvent_IRIS_20141025_145828_3880106953_2014-10-25T14%3A58%3A282014-10-25T14%3A58%3A28.xml

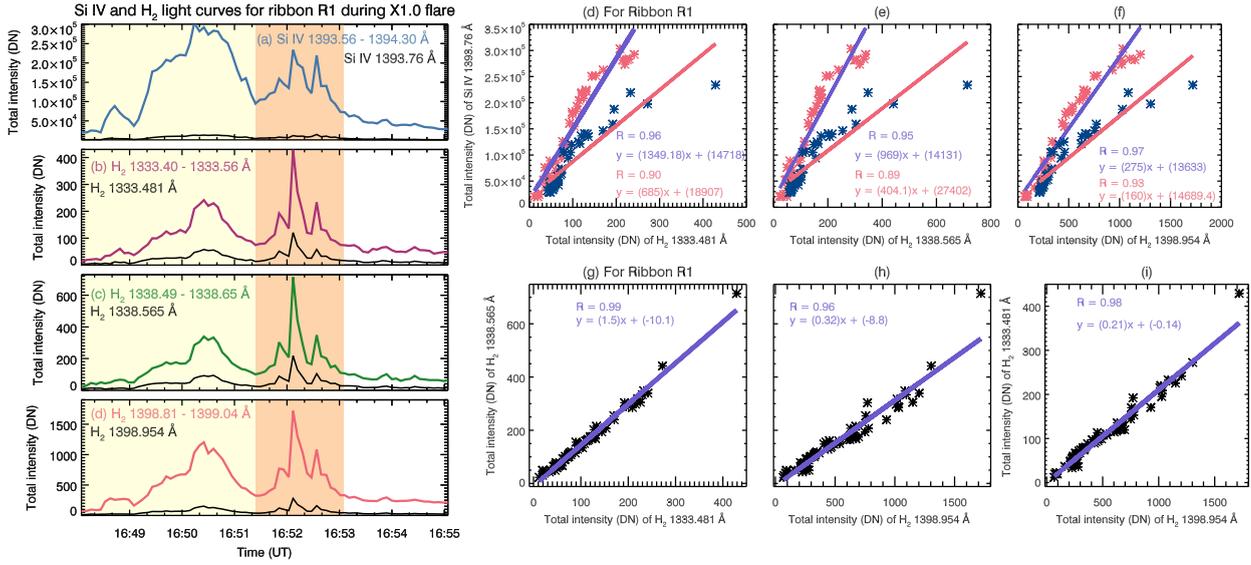

**Figure A1.** Temporal evolution of Si iv and $H_2$ emission at ribbon R1 during the X1.0 flare. Panel (a): Temporal evolution of Si iv (blue curve) (the intensity summed over 1393.56 and 1394.30 Å) and at a single wavelength value of Si iv 1394.76 Å (black curve). Panel (b–d): The temporal evolution of three $H_2$ lines (originated from the same upper level, $(v', j') = (0, 1)$) at a single wavelength value (black curve) and summed over a range of wavelengths (purple, green and pink curves). The intensity variation was observed for two-time slots, and they are highlighted with yellow and orange backgrounds. Panels (d–f): Three scatter plots for Si iv (1393.56 and 1394.30 Å), and three $H_2$ lines (summed over intensities between 1333.40 and 1333.56 Å, between 1338.49 and 1338.65 Å, and between 1398.81 and 1399.04 Å). The intensities are displayed with pink, and blue star symbols. The blue (pink) solid line indicates the linear fit to the data shown in the yellow (orange) background. Panels (g–i): The scatter plots for three $H_2$ lines and indicate the correlation between $H_2$ lines. The intensities are displayed with black star symbols, and solid blue lines indicate the linear fit to the data. The equations for the fitted lines along with fit parameters are given and the Pearson correlation coefficients are displayed as 'R'.